\documentclass[preprint,showpacs,preprintnumbers,amsmath,amssymb]{revtex4}

\usepackage{graphicx}
\usepackage{dcolumn}
\usepackage{color} 
\usepackage{bm}

\usepackage[colorlinks]{hyperref}
\hypersetup{
	pdfstartview={FitH},
	linkcolor=blue,
	citecolor=blue,
	filecolor=blue,
	urlcolor=blue
}

\begin{document}

\title{Universal Quasi-Static Limit for Plasmon-Enhanced Optical Chirality}

\author{Marco Finazzi}
\email{marco.finazzi@polimi.it}
\affiliation{Dipartimento di Fisica and CNISM, Politecnico di Milano,
Piazza Leonardo da Vinci 32, 20133 Milano, Italy}
\author{Paolo Biagioni} \author{Michele Celebrano} \author{Lamberto Du\`{o}}
\affiliation{Dipartimento di Fisica and CNISM, Politecnico di Milano,
Piazza Leonardo da Vinci 32, 20133 Milano, Italy}
\date{\today}

\begin{abstract}
We discuss the possibility of enhancing the chiroptical response from molecules uniformly distributed around nanostructures that sustain localized plasmon resonances.
We demonstrate that the average optical chirality in the near field of any plasmonic nanostrucure cannot be significantly higher than that in a plane wave. This conclusion stems from the quasi-static nature of the nanoparticle-enhanced electromagnetic fields and from the fact that, at optical frequencies, the magnetic response of matter is much weaker than the electric one.
\end{abstract}

\pacs{42.25.Ja,33.55.+b, 78.20.Ek}

\maketitle

Phenomena involving spontaneous symmetry breaking or chiral organization are of fundamental interest in physics and chemistry, and have important implications for biology and life sciences. An example of primary importance is represented by the ubiquitous and yet unexplained homochirality of all biosystems, which are composed by molecules such as sugars and amino-acids that display only one type of chirality \cite{Meierhenrich}. The crucial consequence is that the stereoscopic conformation of a molecule can significantly influence its interaction with the metabolism of an organism \cite{Eriksson}, thus justifying increasing efforts from the scientific community to design new tools for capturing information concerning the chiral properties of organic materials.

Several spectroscopy \cite{Barron, Berova} and microscopy \cite{Savoini2011, Savoini2012} techniques have been developed to this purpose. Such methods typically measure small differences (dichroism) in the interaction of left- and right-circularly polarized light with a chiral material. However, the inherent weakness of the chiroptical properties of organic materials usually hinders the application of such techniques to small amounts of matter. Nevertheless, it has been recently proposed \cite{Tang 2010} and experimentally demonstrated \cite{Tang 2011} that one can create electromagnetic fields displaying  a larger chiral asymmetry compared with circularly polarized plane light waves. Such so-called superchiral fields thus allow measuring the chiroptical properties of small amounts of matter with a large gain in sensitivity \cite{Tang 2011,Choi}.

Another proposed approach to increase chiroptical sensitivity is to exploit engineered light-matter interactions to obtain large field enhancements in extremely limited detection volumes. For instance, resonant nanoantennas are able to produce intense local fields with controlled optical chirality \cite{Biagioni PRL, Biagioni PRB, Lin}. Nanostructures supporting optical resonances have been proven to be able to enhance the circular dichroism spectrum of chiral molecules interacting with them \cite{Govorov, Extarri, Maoz, Lu, Zhao}. It is also possible to engineer resonant nanostructures \cite{Hendry, Davis} or metasurfaces \cite{Yoo, Valev} to generate superchiral local fields.

Although few seminal works already critically addressed the possibilities opened by nanoparticles in the field of optically-enhanced chirality \cite{Choi,Extarri,Schaferling}, the intrinsic limitations of this approach have not yet been fully discussed. In the following we will rigorously demonstrate that the average optically-enhanced chirality in the proximity of nanostructures is subject to a universal limit that arises from the quasi-static nature of the nanoparticle-enhanced electromagnetic fields and from the fact that, at optical frequencies, the magnetic response of metals able to sustain plasma oscillations is much weaker than the electric one.

\begin{figure}
\includegraphics[width=0.4\columnwidth]{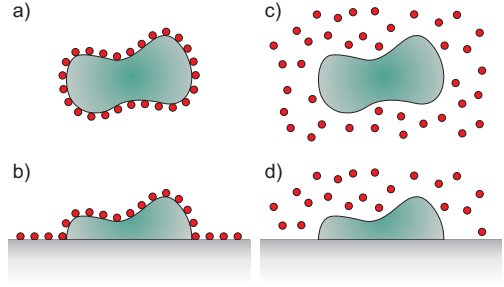}
\caption{\label{fig:1} (Color online) Uniform distribution of chiral molecules (red dots) on the surface [(a) and (b)] or in the volume [(c) and (d)] surrounding a nanoparticle suspended in a fluid [(a) and (c)] or deposited on a substrate [(b) and (d)].}
\end{figure}

We will evaluate the realistic case consisting in a homogeneous distribution of chiral molecules in a domain $\mathcal{D}$ encircling a nanoparticle. $\mathcal{D}$ might coincide with the particle surface in the case the latter has been functionalized with organic surfactants binding the molecules, as in Figs.~\ref{fig:1}(a) and \ref{fig:1}(b), or with the volume around the particle, as in Figs.~\ref{fig:1}(c) and \ref{fig:1}(d) describing the situation where the molecules are suspended in the surrounding liquid.

The signal-to-noise ratio in a dichroic absorption experiment is proportional to $g^{2}I$ \cite{Funk}, with $I$ being the intensity of the light which is modulated between the two opposite circular polarizations and $g$ the dissymmetry factor. The latter is defined as $g=2\left(A^+ - A^-\right)/\left(A^+ + A^-\right)$, where $A^\pm$ is the absorption rate for the two electromagnetic fields interchanged by parity. Tang and Cohen \cite{Tang 2010} demonstrated that, for randomly oriented molecules subject to electric and magnetic dipole transitions at frequency $\omega$,
\begin{equation}\label{eq:1}
    g  = -\left(\frac{G''}{\alpha''}\right)\frac{2}{\omega}\frac{C}{U_E},
\end{equation}
where $C$ and $U_E$ are the time-averaged optical chirality and \textit{electric }energy density, respectively. The quantity in round brackets is related only to the physical properties of the chiral molecules and depends on $\alpha''$ and on $G''$, the imaginary part of the electric polarizability and the isotropic mixed electric-magnetic dipole polarizability, respectively.

Since we are interested in evaluating the dichroic signal from the ensemble of chiral molecules, we need to determine $C$ by averaging over the homogeneous domain $\mathcal{D}$ that contains the molecules. We consider harmonic electromagnetic fields, which can be expressed as follows:
\begin{equation}\label{eq:2}
    \mathbf{E}(\mathbf{r},t) = \hat{\mathbf{E}}(\mathbf{r})e^{-i\omega t}, \mathbf{B}(\mathbf{r},t) = \hat{\mathbf{B}}(\mathbf{r})e^{-i\omega t},
\end{equation}
where $\hat{\mathbf{E}}(\mathbf{r})$ and $\hat{\mathbf{B}}(\mathbf{r})$ are complex vectors: only the real parts of $\mathbf{E}(\mathbf{r},t)$ and $\mathbf{B}(\mathbf{r},t)$ describe the physical electromagnetic fields. Following Ref.~\onlinecite{Tang 2010} we can then write
\begin{equation}\label{eq:3}
C = -\frac{1}{\mathcal{V}_\mathcal{D}}\frac{\epsilon \omega}{2}\mathrm{Im}\left[\int_\mathcal{D} \hat{\mathbf{E}}^*(\mathbf{r})\cdot \hat{\mathbf{B}}(\mathbf{r}) d\mathbf{r}\right],
\end{equation}
$\epsilon$ being the permittivity of the medium surrounding the particle and $\mathcal{V}_\mathcal{D}$ standing for the area [see Figs.~\ref{fig:1}(a) and \ref{fig:1}(b)] or volume [Figs.~\ref{fig:1}(c) and \ref{fig:1}(d)] of the domain $\mathcal{D}$.

For the same reason, in Eq.~(\ref{eq:1}) we need to consider the $\mathcal{D}$-averaged $U_E$ value, equal to
\begin{equation}\label{eq:4}
U_E = \frac{1}{\mathcal{V}_\mathcal{D}}\frac{\epsilon}{4}\int_\mathcal{D} |\hat{\mathbf{E}}(\mathbf{\mathbf{r}})|^2 d\mathbf{r}.
\end{equation}
By applying the Cauchy-Schwarz inequality it is possible to establish a limiting upper value for $|g|$:
\begin{eqnarray}\label{eq:5}
|g| &\le& \left|\frac{G''}{\alpha''} \right|\frac{\epsilon}{\mathcal{V}_\mathcal{D}} \frac{\left|\int_\mathcal{D} \hat{\mathbf{E}}^*(\mathbf{r})\cdot \hat{\mathbf{B}}(\mathbf{r}) d\mathbf{r}\right|}{U_E}\nonumber\\
&\le& \left|\frac{G''}{\alpha''}\right|\frac{4}{c}\frac{\left[\frac{1}{\mathcal{V}_\mathcal{D}}\frac{\epsilon}{4}\int_\mathcal{D} |\hat{\mathbf{E}}(\mathbf{\mathbf{r}})|^2 d\mathbf{r}\right]^\frac{1}{2} \left[\frac{1}{\mathcal{V}_\mathcal{D}}\frac{1}{4\mu}\int_\mathcal{D} |\hat{\mathbf{B}}(\mathbf{\mathbf{r}})|^2 d\mathbf{r}\right]^\frac{1}{2}}{U_E}\nonumber\\
&=& \left|\frac{G''}{\alpha''}\right| \frac{4}{c} \sqrt{\frac{U_B}{U_E}} = |g_{\mathrm{max}}|.
\end{eqnarray}
In this inequality $\mu$ and $c$ are the permeability and the speed of light in the medium hosting the chiral molecules, while $U_B$ is the total time- and space-averaged \textit{magnetic} energy density in $\mathcal{D}$:
\begin{equation}\label{eq:6}
U_B = \frac{1}{\mathcal{V}_\mathcal{D}} \frac{1}{4\mu}\int_\mathcal{D} |\hat{\mathbf{B}}(\mathbf{\mathbf{r}})|^2 d\mathbf{r}.
\end{equation}

In a plane wave $U_B/U_E=1$, but more favorable field geometries indeed exist. The strategy devised by Tang and Cohen to maximize the $g$ value and obtain superchiral fields characterized by $U_B\gg U_E$ consists in choosing the integration domain $\mathcal{D}$ around a nodal surface of the electric field $\hat{\mathbf{E}}(\mathbf{r})$ in a stationary wave, hence minimizing the value of $U_E$ with respect to $U_B$ \cite{Tang 2010}. In practice this is obtained by reflecting circularly polarized light at normal incidence off a mirror with reflectivity close to unity \cite{Tang 2011}.

A fist possible strategy to translate Tang and Cohen's approach into the plasmonic realm would be to replace the mirror with a metal nanoparticle. The  electric field $\hat{\mathbf{E}}_p$ generated by the charges induced in the particle by the illuminating field $\hat{\mathbf{E}}_0$ can be rather accurately described within the so-called \textit{quasi static approximation} \cite{Novotny Hecht}, which neglects retardation effects. In this case the electric field $\hat{\mathbf{E}}_p(\mathbf{r})$ in the domain $\mathcal{D}$ can be expressed as the gradient of a time-varying electrical potential satisfying the Laplace equation, therefore fulfilling the condition $\nabla\times\hat{\mathbf{E}}_p(\mathbf{r})=0$, while the curl of the illumination field is given by $\nabla\times\hat{\mathbf{E}}_0(\mathbf{r}) = \partial\hat{\mathbf{B}}_0(\mathbf{r})/\partial t$. As a consequence, $\hat{\mathbf{E}}_p$ and $\hat{\mathbf{E}}_0$ cannot cancel each other on a connected surface wrapping the particle because, according to the Kelvin-Stokes theorem, they must have different line integrals over any closed loop \cite{Jackson}.

Note that this argument does not exclude that $\hat{\mathbf{E}}(\mathbf{r}) = \hat{\mathbf{E}}_p(\mathbf{r}) + \hat{\mathbf{E}}_0(\mathbf{r}) \approx 0$ for a discrete set of points in space, which would lead to electromagnetic fields that are \emph{locally} superchiral at selected positions in the proximity of a nanoparticle \cite{Schaferling}. However, the impossibility to cancel the impinging electric field with the one scattered by the particle leads to the conclusion that the only way to realize a large \emph{average} optical chirality in the near field of an optical nanoantenna would be to engineer geometries able to generate much larger magnetic than electric near fields. In this way one might fulfill the required condition $U_B \gg U_E$ (with $U_E\neq0$) to obtain $|g_{\mathrm{max}}|$ values significantly higher than those expected for dichroic experiments performed with plane-wave illumination. Since the magnetic response of a non magnetic nanoparticle can be sizeable only when the wavelength of the illuminating light is tuned at one of the particle resonances \cite{Merlin}, one should look for particle geometries sustaining quasi-normal modes characterized by large magnetic multipole moments as compared with the corresponding electric ones. Since plasmon resonances are characterized by high local-field intensity enhancements, in the following we will neglect any interference between the particle quasi-static near field and the illuminating light.

With plane-wave illumination at wavelength $\lambda$, the probability of inducing an electric or magnetic multiplole moment in a particle of size $d<\lambda$ located at \textbf{r} = 0 rapidly falls off with multipole order \cite{Jackson}. For nonmagnetic particles (i.e. particles with vanishing magnetization), the leading-order (in $d/\lambda$) contributions are given by the following expressions \cite{Barron}:
\begin{subequations}\label{eq:7}
\begin{equation}\label{subeq:7a}
    p_i = \underset{j,k}{\sum} \alpha_{ij}\hat{E}_j(0)+\frac{1}{3}A_{ijk}\frac{\partial\hat{E}_j}{\partial x_k}(0) + G_{ij}\hat{B}_j(0) + \ldots
\end{equation}
\begin{equation}\label{subeq:7b}
    m_i = \underset{j}{\sum} \gamma_{ij}\hat{B}_j(0) + G^*_{ij}\hat{E}_j(0) + \ldots
\end{equation}
\begin{equation}\label{subeq:7c}
    q_{ij} = \underset{k}{\sum} A^*_{ijk}\hat{E}_k(0) + \ldots
\end{equation}
\end{subequations}
In these equations, $p_i$, $m_i$ and $q_{ij}$ are the electric dipole, magnetic dipole and electric quadrupole, respectively, while $\alpha_{ij}$ and $\gamma_{ij}$ are the particle polarizability and magnetizability tensors. $G_{ij}$ is the mixed electric-magnetic dipole polarizability of the particle, while $A_{ijk}$ is a third-order tensor that does not contribute to the particle magnetic moment.

At this point, it is worth to remind that, in atoms and molecules, $\alpha_{ij}$, $\gamma_{ij}$,and $G_{ij}$ follow a well-defined hierarchy: $|\alpha_{ij}/G_{i'j'}|\approx\lambda/d_\mathrm{m}$ and $|\alpha_{ij}/\gamma_{i'j'}|\approx(\lambda/d_\mathrm{m})^2$, with $d_\mathrm{m}$ being the molecule (or atom) size \cite{Merlin}. Nanoparticles, however, do not necessarily need to follow this scaling rule. Magnetic multipoles comparatively larger than the corresponding electric ones might in principle be obtained with a proper choice of the nanoantenna geometry and material. However, in metals, this is possible only in the small skin-depth limit associated with the condition $\kappa \gg \lambda/(2\pi d)$ \cite{Merlin}, where $\kappa$ is the imaginary part of the complex refraction index of the particle material, $\lambda/(2\pi d)$ typically being of the order of unity for Ag or Au nanoparticles sustaining plasmonic resonances in the visible. Unfortunately, such condition is not really met for these materials, where $2 \lesssim \kappa \lesssim 5$ for $400 \leq \lambda \leq$ 800~nm \cite{Johnson}. We can thus exclude that a space-integrated optical chirality significantly larger than that of a circularly polarized plane wave could ever be expected in the near field of any plasmonic nanostructure displaying a resonance in the visible region of the electromagnetic spectrum.

\begin{figure}
\includegraphics[width=0.4\columnwidth]{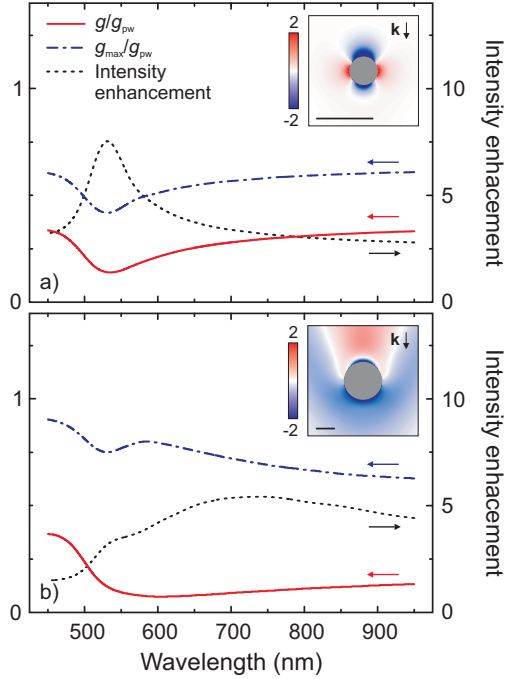}
\caption{\label{fig:2} (Color online) Spectral dependence of the $g/g_{\mathrm{pw}}$ (full lines) and $g_{\mathrm{max}}/g_{\mathrm{pw}}$ (dash-dotted lines) for Au nanospheres with diameter $D$ = 50~nm (a) and 200~nm (b). The respective electric field intensity enhancement spectra are also reported (dotted lines). The insets display the space distribution of the ratio $\left(C-C_{\mathrm{pw}}\right)/C_{\mathrm{pw}}$ ($C_{\mathrm{pw}}$ being the optical chirality of the plane wave) in the sphere symmetry plane \textit{parallel} to the light wavevector \textbf{k}, evaluated at the maximum near field enhancement. The black bar in the insets represents a length of 100~nm.}
\end{figure}

\begin{figure}
\includegraphics[width=0.4\columnwidth]{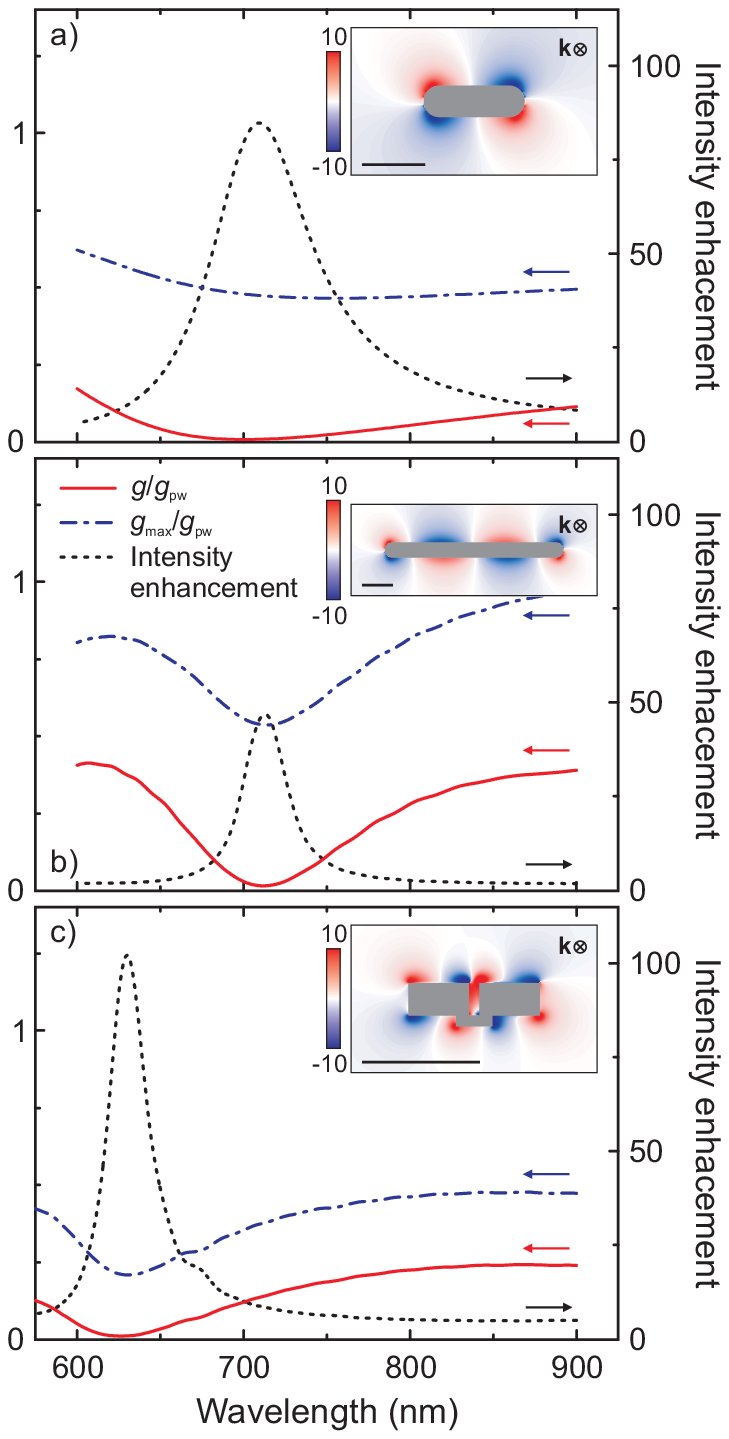}
\caption{\label{fig:3} (Color online) Spectral dependence of the $g/g_{\mathrm{pw}}$ (full lines) and $g_{\mathrm{max}}/g_{\mathrm{pw}}$ (dash-dotted lines) for free-standing Au nanostructures. (a,b) Rounded nanorods with section diameter $D$ = 50~nm and total length $L=160$~nm (a) and 580~nm (b). The simulation has been performed by orienting the rod long axis perpendicular to the light propagation direction. (c) Split-ring antenna with the same geometry as in Ref.~\onlinecite{Feichtner}, simulated by keeping the light wavevector perpendicular to the plane of the structure. The electric field intensity enhancement spectra are also reported (dotted lines). The insets display the space distribution of the ratio $\left(C-C_{\mathrm{pw}}\right)/C_{\mathrm{pw}}$ in the nanostructure symmetry plane \textit{perpendicular} to the light wavevector \textbf{k}, evaluated at the near-field resonance. The black bar in the insets represents a length of 100~nm.}
\end{figure}

To provide further evidence to our thesis and reinforce our conclusions, we have performed numerical simulations with the finite-difference time-domain method \cite{Lumerical} of the electromagnetic fields in the proximity of Au nanospheres, which are widely employed in  many applications of plasmonics and nanosensing. Following the above discussion, spherical noble-metal nanoparticles are ill-suited for solution-phase measurements of the chiroptical properties of molecules, as already pointed out in Ref.~\onlinecite{Extarri}. This is confirmed by Fig.~\ref{fig:2}, which reports the integrated $|g/g_{\mathrm{pw}}|$ value calculated around an Au particles with diameter $D$ = 50 and 200~nm as a function of the excitation wavelength $\lambda$, $g_{\mathrm{pw}}$ being the dissymmetry factor of the same molecules as measured with circularly polarized plane waves, i.e. without the nanoparticle. Figure~\ref{fig:2} also displays $|g_{\mathrm{max}}/g_{\mathrm{pw}}|$ and the electric field intensity enhancement averaged over the spherical surface of the particles \cite{note}. Simulations thus clearly support the conclusion $U_B < U_E$ and, consequently, $|g| < |g_{\mathrm{max}}| < |g_{\mathrm{pw}}|$, also for the sphere with $D =$ 200~nm, for which retardation effects cannot be completely neglected.

The same result has been obtained for a different geometry (see Fig.~\ref{fig:3}a), namely for a gold rod with spherical terminations (cross section diameter $D$ = 50~nm, total length $L$ = 160~nm). Such a geometry leads to a lowest-order resonance located at $\lambda\approx$ 710~nm and to an electric field intensity enhancement of about a factor 85. This higher field-enhancement value compared to the one obtained for the Au nanospheres stems from the fact that the nanorod resonance is located in a spectral region where interband transitions in gold do not occur and the material is thus characterized by much lower losses. Nevertheless, the rod geometry is equally deceiving in terms of optical chirality, as expected from the previous discussion. Noteworthy, a similar result ($|g|<|g_{\mathrm{max}}|<|g_{\mathrm{pw}}|$) is also obtained for a longer rod ($L$ = 580~nm, $D$ = 50~nm), which displays a \textit{third} order resonance of the guided plasmonic mode in the same spectral range (see Fig.~\ref{fig:3}b). This suggests that the conclusions derived above can, to some extent, be prolonged to particles with a larger size than the light wavelength.

The optical behavior of the nanostructures described above is dominated by electric dipole (or multipole) resonances. Split-ring resonators, on the other side, are known to sustain circulating currents with a large associated magnetic dipole \cite{Rockstuhl}. However, it is also known that their large kinetic inductance prevents such resonances from entering the visible spectral range \cite{Zhou}. Recently, Feichtner \textit{et al}. demonstrated that a novel antenna geometry, merging together the characteristics of a dipole and of a split-ring antenna, is able to push the split-ring resonance below 700~nm wavelength \cite{Feichtner}. This geometry, therefore, represents an interesting benchmark to test the limitations outlined above for plasmon-enhanced chirality also in the case non-dipolar resonances. Neverthless, this structure is not able to deliver an averaged optically chirality higher than the one associated to a plane wave, showing a behavior qualitatively similar to the one displayed by the previously discussed geometries (see Fig.~\ref{fig:3}c).

We would like to conclude by noting that, for all the considered geometries, the space-averaged optical chirality of the fields surrounding the particles shows a pronounced minimum at the plasmon resonance. This is a consequence of a large particle polarizability and enhanced electric fields, to be compared to the much smaller magnetic fields resulting from a weak particle magnetizability.

In summary, we have demonstrated that the difficulty of realizing nanostructures capable of  generating near fields with a high average optical chirality is related to fundamental physical limitations to all subwavelength noble-metal particles. This limitation is a consequence of both the quasi-static nature of the nanoparticle-enhanced electromagnetic fields and of the fact that the magnetic response of matter cannot be much higher than the electric one at optical frequencies. This conclusion suggests a twofold strategy to exploit the locally enhanced fields produced by nanostructures to perform high-sensitivity experiments aiming at characterizing the chiroptical properties of molecules. By recalling that the signal-to-noise ratio in a dichroism experiment depends on $g^2 I$ ($I$ being the light intensity), the first possibility would be to selectively place the molecules to be investigated in specific spots where properly engineered particles could concentrate superchiral light ($g \gg 1$). The second solution would be to renounce to superchirality ($g \simeq 1$) and improve the experimental sensitivity by exploiting the enhancement of the light intensity provided by the nanostructure itself.


\begin{thebibliography}{}

\bibitem{Meierhenrich} U. Meierhenrich, \textit{Amino Acids and the Asymmetry of Life} (Springer, New York, 2008).

\bibitem{Eriksson} T. Eriksson, S. Bj\"{o}rkman, B. Roth, {\AA}. Fyge, and P. H\"{o}glund, \href{http://dx.doi.org/10.1002/chir.530070109}{Chirality \textbf{7}, 44 (1995)}.

\bibitem{Barron} L. D. Barron, \textit{Molecular Light Scattering and Optical Activity}
    (Cambridge Univ. Press, Cambridge, 2004).

\bibitem{Berova} N. Berova, P. L. Polavarapu, K. Nakanishi, and W. Woody, editors, \textit{Comprehensive Chiroptical Spectroscopy} Vols. 1, 2 (Wiley, 2012).

\bibitem{Savoini2011} M. Savoini, P. Biagioni, S. C. J. Meskers, L. Du\`{o}, B. Hecht, and M. Finazzi.
    \href{http://pubs.acs.org/doi/abs/10.1021/jz200524m}{J. Phys. Chem. Lett. \textbf{2}, 1359 (2011)}.

\bibitem{Savoini2012} M. Savoini, X. Wu, M. Celebrano, J. Ziegler, P. Biagioni, S. C. J. Meskers, L. Du\`{o}, B. Hecht, and M. Finazzi, \href{http://pubs.acs.org/doi/abs/10.1021/ja209916y}{J. Am. Chem. Soc. \textbf{134}, 5832 (2012)}.

\bibitem{Tang 2010} Y. Tang and A. E. Cohen, \href{http://link.aps.org/doi/10.1103/PhysRevLett.104.163901}{Phys. Rev. Lett. \textbf{104}, 163901 (2010)}.

\bibitem{Tang 2011} Y. Tang and A. E. Cohen, \href{http://www.sciencemag.org/content/332/6027/333.abstract}{Science \textbf{332}, 333 (2011)}.

\bibitem{Choi} J. S. Choi and M. Cho, \href{http://link.aps.org/doi/10.1103/PhysRevA.86.063834}{Phys. Rev. A \textbf{86}, 063834 (2012)}.

\bibitem{Biagioni PRL} P. Biagioni, J.-S. Huang, L. Du\`{o}, M. Finazzi, and B. Hecht, \href{http://link.aps.org/doi/10.1103/PhysRevLett.102.256801}{Phys. Rev. Lett. \textbf{102}, 256801 (2009)}.

\bibitem{Biagioni PRB} P. Biagioni, M. Savoini, J. S. Huang, L. Du\`{o}, M. Finazzi, and B. Hecht,
    \href{http://link.aps.org/doi/10.1103/PhysRevB.80.153409}{Phys. Rev. B \textbf{80}, 153409 (2009)}.

\bibitem{Lin} D. Lin and J.-S. Huang, \href{http://www.opticsexpress.org/abstract.cfm?URI=oe-22-7-7434}{Opt. Express \textbf{22}, 7434 (2014)}.

\bibitem{Govorov} A. O. Govorov, Z. Fan, P. Hernandez, J. M. Slocik, and R. R. Naik,
    \href{http://pubs.acs.org/doi/abs/10.1021/nl100010v}{Nano Lett. \textbf{10}, 1374 (2010)}.

\bibitem{Extarri} A. Garc\'{i}a-Etxarri and J. A. Dionne, \href{http://link.aps.org/doi/10.1103/PhysRevB.87.235409}{Phys. Rev. B \textbf{87}, 235409 (2013)}.

\bibitem{Maoz} B. M. Maoz, Y. Chaikin, A. B. Tesler, O. Bar Elli, Z. Fan, A. O. Govorov, and G. Markovich,
    \href{http://pubs.acs.org/doi/abs/10.1021/nl304638a}{Nano Lett. \textbf{13}, 1203 (2013)}.

\bibitem{Lu} F. Lu, Y. Tian, M. Liu, D. Su, H. Zhang, A. O. Govorov and O. Gang,
    \href{http://pubs.acs.org/doi/abs/10.1021/nl401107g}{Nano Lett. \textbf{13}, 3145 (2013)}.

\bibitem{Zhao} Y. Zhao, L. Xu, W. Ma, L. Wang, H. Kuang, C. Xu, and N. A. Kotov,
    \href{http://dx.doi.org/10.1021/nl501166m}{Nano Lett. \textbf{14}, 3908 (2014)}.

\bibitem{Hendry} E. Hendry, T. Carpy, J. Johnston, M. Popland, R. V. Mikhaylovskiy, A. J. Lapthorn, S. M. Kelly, L. D. Barron, N. Gadegaard, and M. Kadodwala, \href{http://www.nature.com/nnano/journal/v5/n11/abs/nnano.2010.209.html}{Nat. Nanotechnol. \textbf{5}, 783 (2010)}.

\bibitem{Davis} J. T. Davis and E. Hendry, \href{http://link.aps.org/doi/10.1103/PhysRevB.87.085405}{Phys. Rev. B \textbf{87}, 085405 (2013)}.

\bibitem{Yoo} S. J. Yoo, M. Cho, and Q-Han Park,
    \href{http://link.aps.org/doi/10.1103/PhysRevB.89.161405}{Phys. Rev. B \textbf{89}, 161405(R) (2014)}.

\bibitem{Valev} V. K. Valev et al., \href{http://dx.doi.org/10.1002/adma.201401021}{Adv. Mater. \textbf{26}, 4074 (2014)}.

\bibitem{Schaferling} M. Sch\"{a}ferling, D. Dregely, M. Hentschel, and H. Giessen,
    \href{http://link.aps.org/doi/10.1103/PhysRevX.2.031010}{Phys. Rev. X \textbf{2}, 031010 (2012)}.

\bibitem{Funk} T. Funk, A. Deb, S. J. George, H. X. Wang and S. P. Cramer, \href{http://www.sciencedirect.com/science/article/pii/S0010854504001353}{Coord. Chem. Rev. \textbf{249}, 3 (2005)}.

\bibitem{Novotny Hecht} L. Novotny and B. Hecht, \textit{Principles of Nano-Optics} (Cambridge University Press, Cambridge, 2006).

\bibitem{Jackson} J. D. Jackson, \textit{Classical Electrodynamics} (Wiley, 1999).

\bibitem{Merlin} R. Merlin, \href{http://www.pnas.org/content/106/6/1693.abstract}{Proc. Nat. Acad. Sci. \textbf{106}, 1693 (2009)}.

\bibitem{Johnson} P. B. Johnson and R. W. Christy, \href{http://link.aps.org/doi/10.1103/PhysRevB.6.4370}
    {Phys. Rev. B \textbf{6}, 4370 (1972)}.

\bibitem{Lumerical} \textit{FDTD Solutions} v. 6.0.5, Lumerical Solutions Inc., Vancouver, Canada.

\bibitem{note} To avoid numerical instabilities due to the square grid discretization, we have evaluated the spectra reported in Figs.~\ref{fig:2} and \ref{fig:3} by integrating the corresponding quantities over a surface located at a distance of 4~nm from the particle surface.

\bibitem{Rockstuhl} C. Rockstuhl, F. Lederer, C. Etrich, T. Zentgraf, J. Kuhl, and H. Giessen, Opt. Express 14, 8827 (2006)

\bibitem{Zhou} J. Zhou, T. Koschny, M. Kafesaki, E. N. Economou, J. B. Pendry, and C. M. Soukoulis, Phys. Rev. Lett. 95, 223902 (2005)

\bibitem{Feichtner} T. Feichtner, O. Selig, M. Kiunke, and B. Hecht, \href{http://link.aps.org/doi/10.1103/PhysRevLett.109.127701}{Phys. Rev. Lett. \textbf{109}, 127701 (2012)}.


\end{thebibliography}
\end{document}